\documentclass[aps,prb,twocolumn,floatfix]{revtex4-2}
\usepackage{graphicx,graphics}
\usepackage{natbib}
\usepackage{lipsum}
\usepackage{dcolumn}
\usepackage{amsmath,amssymb,amsfonts}
\usepackage{latexsym,verbatim}
\usepackage{bm}
\usepackage{color}
\usepackage{siunitx}
\usepackage[breaklinks=true,colorlinks,citecolor=blue,linkcolor=blue,urlcolor=blue]{hyperref}
\begin{document}
	
\title{Generalized first-principles method to study near-field heat transfer mediated by Coulomb interaction}
\author{Tao Zhu}
\email{phyzht@nus.edu.sg}
\affiliation{Department of Physics, National University of Singapore, Singapore 117551, Republic of Singapore}
\author{Jian-Sheng Wang}
\affiliation{Department of Physics, National University of Singapore, Singapore 117551, Republic of Singapore}
\date{\today}

\begin{abstract}
We present a general microscopic first-principles method to study the Coulomb-interaction-mediated heat transfer in the near field. Using the nonequilibrium Green's function formalism, we derive Caroli formulas for heat transfers between materials with translational invariance. The central physical quantities are the screened Coulomb potential and the spectrum function of polarizability. Within the random phase approximation, we calculate the polarizability using the linear response density functional theory and obtain the screened Coulomb potential from a retarded Dyson equation. We show that the heat transfer mediated by the Coulomb interaction is consistent with that of the $p$-polarized evanescent waves which dominate the heat transfer in the near field. We adopt single-layer graphene as an example to calculate heat transfers between two parallel sheets separated by a vacuum gap $d$. Our results show a saturation of heat flux at the extreme near field which is different from the reported $1/d$ dependence for local response functions. The calculated heat flux is up to $5\times10^4$ times more than the black-body limit, and a $1/d^2$ dependence is shown at large separations. From the spectrum of energy current density, we infer that the near-field enhancement of heat transfer stems from electron transitions around the Fermi energy. With a uniform strain, the heat flux increases for most of the distances while a negative correlation is shown at the moderate field. Our method is valid for inhomogeneous materials in which the macroscopic response function used in conventional theory of fluctuational electrodynamics would fail at the subnanometer scale.

\end{abstract}
\maketitle

{\emph{Introduction.}}
From the 1970s, heat transfer in the near field has attracted extensive interest because of its colossal thermal radiation over the black-body limit \cite{volokitin,zzm,biehs}. When the gap size is smaller than Wien's wavelength, experimental works revealed that the heat flux between two parallel plates can achieve thousands of times more than that being predicted by the Stefan-Boltzmann law \cite{experiment1,experiment2,experiment3}. The most widely recognized theoretical explanation of such near-field effects is given by the fluctuational electrodynamics (FE) of Rytov \cite{rytov1,rytov2} with the further development by Poler and Van Hove \cite{polder}. In the FE theory, the heat flux is generated by thermally driven current fluctuations that follow the equilibrium fluctuation-dissipation theorem \cite{fdt1,fdt2}. Based on macroscopic Maxwell’s equations, the energy current is given by a Landauer-type expression with transmission coefficients that consist of contributions from both propagating and evanescent modes. The great enhancement of thermal radiation in the near field is due to the tunneling of evanescent waves that decay exponentially with the gap size \cite{song,cuevas}.

Recent experiments have approached distances down to the scale of nanometers to angstroms in which extraordinarily large heat current was found \cite{experiment4,experiment5,experiment6,experiment7}. The reported near-field enhancements exceed the value predicted by the FE theory and have been partially explained by mechanisms such as phonon tunneling and potential contaminants \cite{vazrik,xiong,cui}. However, at the scale of subnanometers, the microscopic inhomogeneity of materials induces local field effects (LFEs) \cite{lfe1} that can significantly alter the response properties of materials \cite{lfe2}, and thus, the phenomenological macroscopic response functions used in the FE theory are insufficient to describe heat transfers between inhomogeneous materials in the extreme near field. On the other hand, many efforts have been devoted to developing microscopic theories beyond the FE theory \cite{negf1,negf2,negf3,negf4}. In the extreme near field, the vector potential of current fluctuations is less important, and the scalar potential of the density fluctuations becomes dominant \cite{mahan}. In the quasi-static limit, the scalar potential gives rise to instantaneous Coulomb interactions between electrons, so that thermally driven density fluctuations induce energy transfers in the near field. Using the nonequilibrium Green's function (NEGF) method, a quantum mechanical scalar field theory of heat transfer was proposed in which the transmission coefficient can be written as a Caroli formula \cite{zhang,caroli}. An equivalent formula can be obtained from the Joule heating effect of charge fluctuations due to external electric fields \cite{Yu,equivalent,zhu}.

This Coulomb-interaction-mediated heat transfer in the near field has been studied for some systems \cite{Yu,zhang,jiang,tang,lian}. However, similar to conventional FE studies, a specific theoretical modeling of the response function is needed for different materials to obtain the transmission coefficient. In this Letter, we introduce a generalized parameter-free first-principles method to calculate the near-field heat flux between two parallel plates. Within the random phase approximation (RPA), we obtain the response function from electronic band structures calculated from density functional theory (DFT) \cite{Onida}. The obtained response function can be used in both the NEGF and FE formalism with proper treatment of the LFEs \cite{zhu}. Since the microscopic NEGF theory naturally captures the inhomogeneities of materials, in this work, we mainly follow the NEGF scheme of heat transfers and derive Caroli formulas for systems with translational invariance in which inhomogeneities are given by reciprocal lattice vectors. Moreover, with macroscopic approximation, we show that the NEGF formalism for the Coulomb heat transfer is equivalent to the $p$-polarized evanescent modes of the FE theory in the quasistatic limit. Last, we adopt graphene as an example to introduce the procedures to obtain the heat flux, and this method can easily be applied to other materials with arbitrary thicknesses and possible inhomogeneity.

{\emph{Method.}}
In the NEGF formalism of heat transfer, the energy current per unit area between two bodies is given by a Landauer-like formula
\begin{equation}
	J= \int_{0}^{\infty}\dfrac{d\omega}{2\pi} \hbar\omega \bigl[N_1(\omega)-N_2(\omega)\bigr]T(\omega),
	\label{landauer}
\end{equation}
where $N_\alpha = 1/(e^{\hbar\omega/(k_BT_\alpha)}-1)$ is the Bose distribution function at the temperature $T_\alpha$ for plate $\alpha$. With the local equilibrium approximation, Eq.~(\ref{landauer}) can be derived from the Meir-Wingreen formula \cite{meir1,meir2}, and the transmission coefficient $T(\omega)$ is given by a Caroli formula: \cite{zhang,jiang}
\begin{equation}
    T(\omega)={\rm Tr}\big[D^r\Gamma_1 D^a\Gamma_2\big],
	\label{caroli}
\end{equation}
where superscripts $r$ and $a$ denote retarded and advanced components, respectively. The screened Coulomb potential $D$ follows the Dyson equation \cite{dyson} $D = D^0+D^0\Pi D$ for the scalar-field Green's function. At the quasi-static limit, $D^0$ has only the component of the pure Coulomb potential $v$. In real space, we can write the retarded Dyson equation as
\begin{eqnarray}
	\label{dyson}
	\nonumber D^r({\bf r},{\bf r}^\prime,\omega) = &v&({\bf r},{\bf r}^\prime) + \int d{\bf r}^{\prime\prime}\int d{\bf r}^{\prime\prime\prime} v({\bf r},{\bf r}^{\prime\prime})\\
	&\times&\Pi^r({\bf r}^{\prime\prime},{\bf r}^{\prime\prime\prime},\omega)D^r({\bf r}^{\prime\prime\prime},{\bf r}^\prime,\omega),
\end{eqnarray}
where $\Pi^r$ is the retarded scalar photon self-energy or the polarizability that represents the linear response of the induced charge density to the total potential of the system. Within the RPA \cite{rpa}, the linear-response Kubo's formalism of polarizability is \cite{kubo,polarizability} 
\begin{equation}
	\label{pi}
	\Pi^r({\bf r},{\bf r}^\prime,\omega)=2e^2\sum_{i, j}(f_j-f_i)\dfrac{\psi_j({\bf r}) \psi^*_j({\bf r}^\prime) \psi^*_i({\bf r}) \psi_i({\bf r}^\prime)}{\epsilon_j-\epsilon_i-\hbar\omega-i\eta},
\end{equation}
where $\psi_{i (j)}$ and $\epsilon_{i (j)}$ are volume-normalized independent particle wave-functions and energies of state $i(j)$, respectively. The function $f = (1+e^{\beta(\epsilon-\mu)})^{-1}$ is the Fermi distribution function, with $\beta = 1/k_BT$ and $\mu$ being the chemical potential. The factor of 2 accounts for the spin degeneracy, and $e$ is the electron charge. The damping factor $\eta$ is a small, positive quantity that accounts for the carrier relaxation of long-range scatterers \cite{eta}.

\begin{figure}
	\includegraphics[width=8.6 cm]{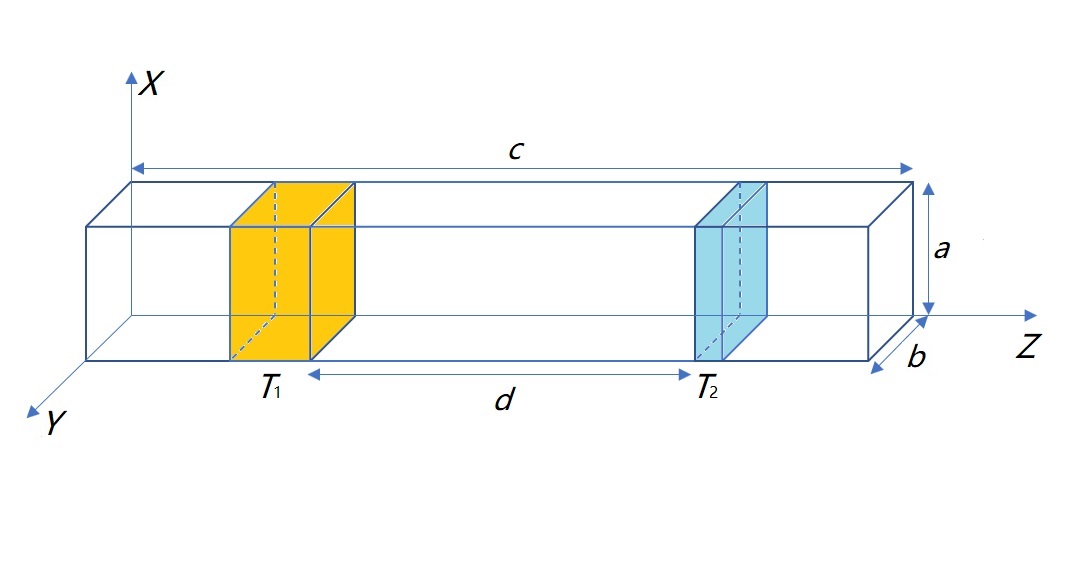}
	\caption{Sketch of the unit simulation box of two parallel plates with a vacuum gap of $d$. Each of the plates is in its own internal thermal equilibrium state; that is, plate 1 has temperature $T_1$, and plate 2 has temperature $T_2$. $a$, $b$, and $c$ are lattice constants of the $x$, $y$, and $z$ directions, respectively.}
	\label{fig1}
\end{figure}
 
In Eq.~(\ref{caroli}), the spectrum function of each plate is defined as $\Gamma_\alpha=i(\Pi^r_\alpha-\Pi_\alpha^a)$, where $\alpha=$ 1 or 2 and $\Pi_\alpha$ is the individual polarizability of plate $\alpha$ in isolation. As the advanced components are the Hermitian transpose of the retarded one, i.e., $D^a=(D^r)^\dagger$ and $\Pi^a=(\Pi^r)^\dagger$, in the following discussion, we omit the superscript and all quantities are retarded unless otherwise specified. 

With Eqs.~(\ref{landauer}) to (\ref{pi}), one can obtain the heat flux between two bodies with arbitrary geometry. However, both polarizability $\Pi$ and screened Coulomb potential $D$ are difficult to solve in real space as the position ${\bf r}$ is continuous. For periodic systems with planar geometry, due to translational invariance $f({\bf r}+{\bf R},{\bf r}^\prime+{\bf R})=f({\bf r},{\bf r}^\prime)$, it is more convenient to calculate these quantities in momentum space with the Bloch wave vector ${\bf q}$ and the reciprocal lattice vector ${\bf G}$. 

We show in Fig.~\ref{fig1} the unit simulation box of two parallel plates separated by a vacuum gap of $d$. From the Dyson equation, the screened Coulomb potential can be written as $D=\varepsilon^{-1}v$, where $\varepsilon$ is the dielectric function that is given by
\begin{equation}
	\varepsilon_{{\bf G},{\bf G}^\prime}({\bf q},\omega)=\delta_{{\bf G},{\bf G}^\prime}-v_{{\bf G},{\bf G}}({\bf q})\Pi_{{\bf G},{\bf G}^\prime}({\bf q},\omega).
	\label{dielectric}
\end{equation}
The Fourier transform of the pure Coulomb potential in momentum space is
\begin{equation}
	v_{{\bf G},{\bf G}^{\prime}}({\bf q}) = \delta_{{\bf G},{\bf G}^\prime}\dfrac{1}{\varepsilon_0\vert{\bf q}+{\bf G}\vert^2},
	\label{coulomb3d}
\end{equation}
where $\varepsilon_0 \approx 8.85\times10^{-12} \,$ F/m is the vacuum permittivity.
 
In RPA, the polarizability can be obtained from DFT with the Adler-Wiser formula \cite{adler,wiser}
\begin{eqnarray}
	\label{pi2}
	&&\Pi_{{\bf G},{\bf G}^\prime}({\bf q},\omega)=\dfrac{2e^2}{\Omega}\sum_{n,n',{\bf k}}\textit{w}_{{\bf k}}\bigl(f_{n'{\bf k}+{\bf q}}-f_{n{\bf k}}\bigr)\\
	&&\nonumber\times\dfrac{\langle\phi_{n{\bf k}}|e^{-i({\bf q}+{\bf G})\cdot {\bf r}}|\phi_{n'{\bf k}+{\bf q}}\rangle\langle\phi_{n'{\bf k}+{\bf q}}|e^{i({\bf q}+{\bf G}^\prime)\cdot  {\bf r}}|\phi_{n{\bf k}}\rangle}{\epsilon_{n'{\bf k}+{\bf q}}-\epsilon_{n{\bf k}}-\hbar \omega-i\eta},
\end{eqnarray}
where $\phi_{n{\bf k}}$ and $\epsilon_{n{\bf k}}$ are Kohn-Sham \cite{ks} eigenfunctions and eigenvalues, respectively. The Fermi occupation function $f$ equals 1 for occupied states and 0 for unoccupied states. $\Omega$ is the volume of the primitive cell, and $\textit{w}_{{\bf k}}$ is the weight of each $k$-point in the first Brillouin zone which accounts for symmetry.

In the momentum space, the transmission coefficient for parallel plates is
\begin{equation}
	\label{caroli2}
	T(\omega)=\frac{1}{A}\sum_{\bf q}{\rm Tr}_{\bf G}\big[D\Gamma_1 D^\dagger\Gamma_2\big],
\end{equation}
where $A$ is the area of plates and ${\bf q}$ lies in the first Brillouin zone. All quantities inside the brackets in Eq.~(\ref{caroli2}) involve matrix multiplications with dimensions of $N\times N$, where $N$ is the number of ${\bf G}$ vectors which is subject to the kinetic energy cutoff of response functions, i.e., $|{\bf q}+{\bf G}|^2<E_{\rm cut}$. The trace is applied to only the ${\bf G}$ space that accounts for microscopic inhomogeneity. 

With Eqs.~(\ref{dielectric}) to (\ref{caroli2}), one can calculate the transmission coefficient of two parallel plates with arbitrary thickness. Nevertheless, because the two plates are localized in the $z$-direction with a vacuum gap of $d$, a sufficiently large number of $G_z$ components is required. If we assume there is no electronic tunneling between the two plates, i.e., the Hamiltonian is block diagonal, the polarizability of the system can be written as $\Pi = \Pi_1+\Pi_2$, which in ${\bf G}$ space is
\begin{equation}
	\begin{array}{ll}
		\Pi=
		\begin{pmatrix}
			\Pi_{11} & 0\\
			0 & \Pi_{22}
		\end{pmatrix},
	\end{array} 
\end{equation}
where $\Pi_{11}$ and $\Pi_{22}$ are submatrices of $\Pi_{1}$ and $\Pi_{2}$ with dimensions of $N_1\times N_1$ and $N_2\times N_2$, respectively. In other words, each diagonal block of the whole polarizability matrix consists of localized polarizability of each plate in isolation and all entries in the off-diagonal block are zero. In this case, we can simplify our problem by separating the whole system into two independent pieces and perform separate calculations of $\Pi_1$ and $\Pi_2$ with the same unit cell and same sets of ${\bf G}$ vectors \cite{quek}. 

Moreover, if both plates are two-dimensional (2D) materials, then electrons are confined in the plane perpendicular to the $z$-axis, i.e., electron density $\rho({\bf r}_\perp,z)\propto\delta(z)$. We can further simplify the problem by introducing the 2D surface density
\begin{equation}
	\sigma({\bf r}_\perp)=\int_{-\infty}^{\infty}\rho({\bf r}_\perp,z)dz
\end{equation}
into Kubo's formalism and define the 2D polarizability as \cite{2dpi}
\begin{equation}
	\label{2D_pi}
	\Pi^{2D}_{{\bf G}_\perp,{\bf G}^\prime_\perp}({\bf q}_\perp,\omega)=c\times\Pi^{3D}_{({\bf G}_\perp,0),({\bf G}^\prime_\perp,0)}\left(({\bf q}_\perp,0),\omega\right),
\end{equation}
where $c$ is the lattice constant in the $z$ direction. Equation~(\ref{2D_pi}) implies that we can reduce the original three-dimensional (3D) problem to a 2D problem by sampling ${\bf q}$ and ${\bf G}$ with only in-plane components, i.e., ${\bf q}=({\bf q}_\perp,0)$ and ${\bf G}=({\bf G}_\perp,0)$. Without cross correlation between two sides, the Dyson equation in Eq.~(\ref{dyson}) has a matrix form, 

\begin{equation}
	\label{dyson2}
	\begin{array}{ll}
		\begin{pmatrix}
			\varepsilon_1 & -v_{12}\Pi_{22}\\
			-v_{21}\Pi_{11} & \varepsilon_2
		\end{pmatrix}
		\begin{pmatrix}
			D_{11} & D_{12}\\
			D_{21} & D_{22}
		\end{pmatrix}
		=
		\begin{pmatrix}
			v_{11} & v_{12}\\
			v_{21} & v_{22}
		\end{pmatrix},
	\end{array} 
\end{equation}
where $\varepsilon_\alpha=I-v_{\alpha\alpha}\Pi_{\alpha\alpha}$ is the dielectric function of plate $\alpha$. The components of the 2D pure Coulomb potential are defined as 
\begin{equation}
	[v_{\alpha\beta}]_{{\bf G},{\bf G^\prime}} = \delta_{{\bf G},{\bf G}^\prime}\dfrac{e^{-\vert {\bf q}+{\bf G}\vert\vert z_\alpha - z_\beta\vert}}{2\varepsilon_0\vert{\bf q}+{\bf G}\vert},
	\label{coulomb2d}
\end{equation}
where $z_{\alpha (\beta)}$ is the coordinate of plate $\alpha$ ($\beta$) in the $z$-axis. Taking the trace of each side explicitly, the transmission coefficient in Eq.~(\ref{caroli2}) can be simplified as
\begin{equation}
	\label{caroli3}
	T(\omega)=\frac{1}{A}\sum_{\bf q}{\rm Tr}_{\bf G}\big[D_{21}\Gamma_{11} D_{12}^\dagger\Gamma_{22}\big],
\end{equation}
where all quantities are in the 2D form and ${\bf q}$ lies in the 2D first Brillouin zone.

It is worth mentioning that Eq.~(\ref{caroli2}) is valid for both 2D and 3D problems, provided that all quantities are consistent in dimension. However, it requires a very large sampling of ${\bf q}$ and ${\bf G}$ to capture the locality of each plate in the $z$ direction, which might limit its practical applications. On the other hand, the computational cost is significantly reduced for the 2D cases with Eqs.~(\ref{2D_pi}) to (\ref{caroli3}) because only in-plane components of ${\bf q}$ and ${\bf G}$ have to be considered. 
	
Furthermore, if the 2D material is homogeneous along both the $x$ and $y$ directions, one can also neglect the LFEs, i.e., set $G_x=G_y=0$. In this case, the difference between microscopic and macroscopic response functions vanishes, so the microscopic NEGF Caroli formula (\ref{caroli3}) is expected to be consistent with the macroscopic FE theory. Actually, from Eq.~(\ref{coulomb2d}), we can introduce $v_q = 1/(2\varepsilon_0q)$ such that $v_{11} = v_{22} = v_q$ and $v_{12} = v_{21} = v_qe^{-qd}$, where $q = |{\bf q}|$ is the magnitude of wave vector. Then, the components of the screened Coulomb potential in Eq.~(\ref{dyson2}) can be written explicitly as
\begin{equation}
	D_{12} = D_{21} = \dfrac{v_qe^{-qd}}{\varepsilon_1\varepsilon_2-e^{-2qd}v_q\Pi_{11}v_q\Pi_{22}}.
\end{equation}
Additionally, the reflection coefficient for TM polarization is $r=(1-\varepsilon)/\varepsilon$ \cite{ilic}. With the spectrum function $\Gamma=i(\Pi-\Pi^\dagger) = -2{\rm Im}(\Pi)$ and a straightforward derivation, we finally have
\begin{equation}
	\label{evanescent}
	T(\omega)=\frac{1}{A}\sum_{\bf q}\dfrac{4{\rm Im}(r_1){\rm Im}(r_2)e^{-2qd}}{\left|1-r_1r_2e^{-2qd}\right|^2},
\end{equation}
which is exactly the FE evanescent modes of the transmission coefficient in the quasistatic limit \cite{ilic}.

{\emph{Computational details.}}
As an example, we calculate the heat flux between two parallel single-layer graphene sheets that have been studied theoretically in the framework of both the NEGF \cite{jiang} and FE theory \cite{ilic,ilic2,pablo} with specific phenomenological treatments of the response function. To obtain the heat flux from the first-principles method, we start from the ground state calculations of graphene by using DFT as implemented in QUANTUM ESPRESSO \cite{qe1,qe2}. We adopt the norm-conserving pseudopotential generated by the Martins-Troullier method \cite{martin} with the Perdew-Burke-Ernzerhof exchange-correlation functional \cite{pbe} in the generalized gradient approximation. The plane-wave basis set with a 60 Ry energy cut-off is used to expand the Kohn-Sham wave functions. Fermi-Dirac smearing with a 0.002 Ry smearing width is employed to treat the partial occupancies. The in-plane lattice constants are $a = b = $ \SI{2.46}{\angstrom}. To avoid interactions from the neighboring lattice in the $z$ direction, a large lattice constant of $c = \SI{18}{\angstrom}$ is set in the $z$ direction of the unit cell.

The polarizability of each side is calculated on top of the ground state band structure by using the BERKERELYGW package \cite{bgw1,bgw2}. A $90\times90\times1$ Monkhorst-Pack \cite{mp} grid is used to sample the first Brillouin zone for the nonlocal polarizability, while the long-wave ($q\to0$) polarizability is obtained from a much finer $300\times300\times1$ grid. To avoid divergence of the Coulomb potential, we use a small value of $q=10^{-5}$ a.u. in the calculation of contributions from the long-wave polarizability. The energy broadening factor $\eta$ is set to \SI{0.05}{\electronvolt}, which corresponds to an electron relaxation time of $10^{-14}$ s \cite{ilic}. We also neglect the LFEs that are important only for systems with inhomogeneous geometry \cite{lfe2}. Then we solve the Dyson equation to get the screened Coulomb potential and calculate the transmission coefficient from the Caroli formula. Last, we integrate over frequencies to get the heat flux. To compare our results with previous reports, we also do a parallel calculation using a local response function of graphene in the Dirac model \cite{polarizability,ilic}.

 \begin{figure}
	\includegraphics[width=8.6 cm]{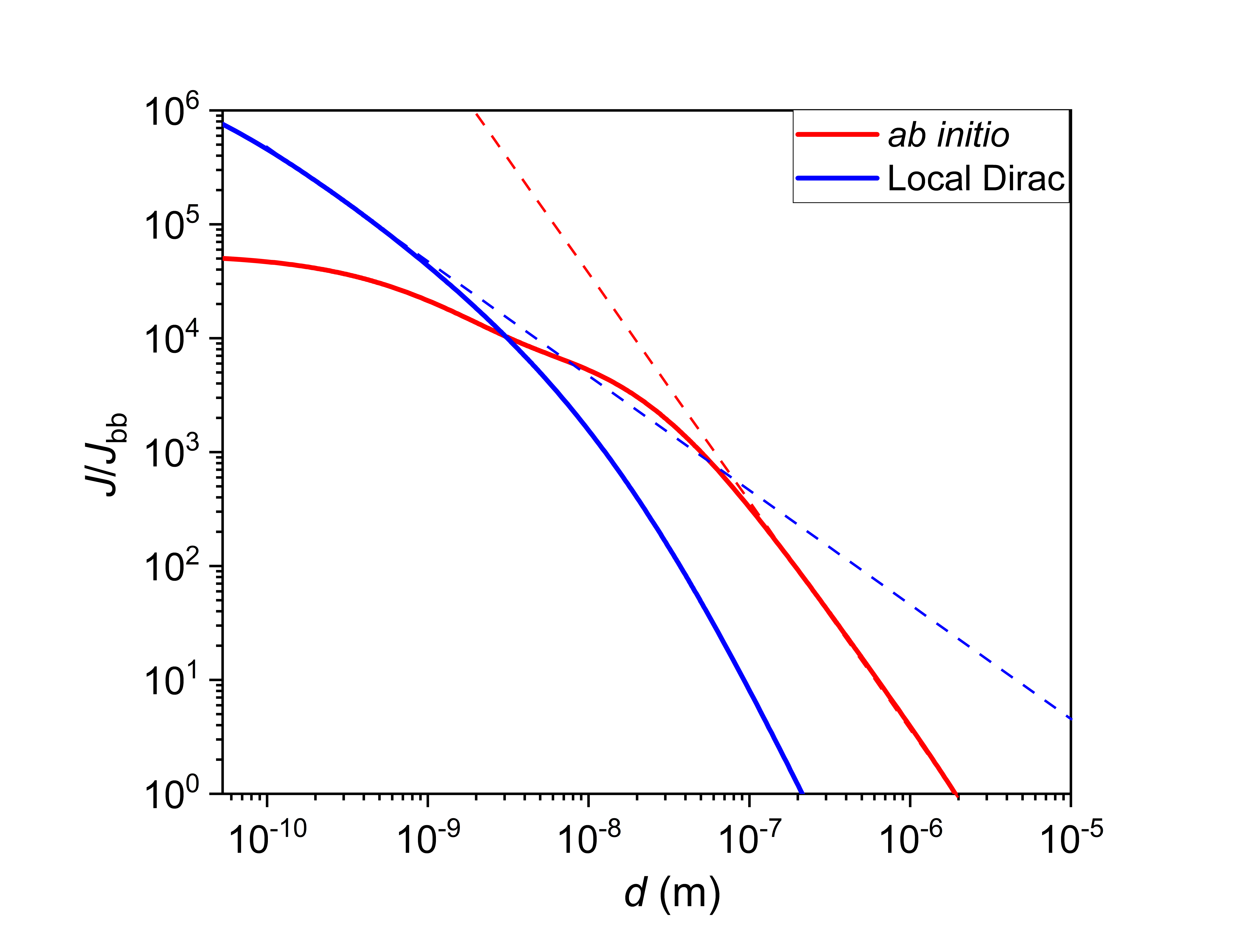}
	\caption{Distance dependence of the near-field heat flux ratio between two parallel graphene sheets. The temperature is fixed at $T_1=$ 1000 K and $T_2=$ 300 K. The red solid curve shows results obtained from \textit{ab initio} first-principles calculation, and the blue solid curve represents results obtained from a local response function of graphene in the Dirac model. The red and blue dashed lines show the asymptotic $1/d^2$ and $1/d$ dependence of heat flux at small and large separations, respectively.}
	\label{fig2}
\end{figure}

{\emph{Results and discussion.}}
In Fig.~\ref{fig2}, we show the calculated heat flux of two parallel graphene sheets as a function of gap sizes. The vertical coordinate is the ratio of the calculated near-field heat flux to the black-body radiative heat flux given by the Stefan-Boltzmann law $J_{bb}=\sigma(T_1^4-T_2^4)$, with $\sigma \approx 5.67\times10^{-8} \,$W/(m$^2$ K$^4$). As shown in Fig.~\ref{fig2}, the near-field heat flux is remarkably larger than that of the black-body radiation. At small separation, a converged ratio around $5\times10^4$ is shown for the \textit{ab initio} results. This value agrees well with a previous report that used the tight-binding method to calculate the density response function of graphene \cite{jiang}. The saturation of heat flux in the extreme near field originates from the nonlocal effect of wave vectors, which is a typical behavior of thermal radiation mediated by $p$-polarized evanescent waves \cite{nonlocal}. Without spatial dispersion, the heat flux calculated from the local response function shows a $1/d$ dependence at short separation, which agrees with the previous report \cite{pablo}.

With an increase in distances, the heat flux decreases monotonically. Due to the exponential factor that appears in the 2D Coulomb potential, the long-wave ($q\to0$) contribution becomes dominant at large distances. When $d>\SI{100}{\nano\meter}$, the \textit{ab initio} heat flux shows an asymptotic dependence of $1/d^2$, which is consistent with near-field heat flux between parallel plate capacitors \cite{negf2}. On the other hand, the heat flux calculated from the local response function decays faster than that from the \textit{ab initio} results. Different power laws of heat flux at large separations have also been reported \cite{jiang,pablo}. According to Eqs.~(\ref{caroli3}) to (\ref{evanescent}), for homogeneous materials like graphene, the NEGF Caroli formula is equivalent to the FE formalism for heat transfers, so all discrepancies between previous reports and our first-principles calculation originate from the different theoretical treatments of the response function. In Ref.~\cite{polarizability}, we gave a more detailed comparison of the response function of graphene obtained from different theoretical approaches. In addition, the uncertainty of numerical errors from finite samplings of the first Brillouin zone may also play a role because exact results require an infinitely large $k$-point sampling. Despite several different power laws that have been reported, we believe the first-principles calculation is the most accurate as the response function is obtained from full DFT band structures. This is further supported by recent experimental work showing that the near-field heat flux between graphene sheets is higher than that predicted by FE with the Dirac response function \cite{grapheneexp}. When $d>\SI{2}{\micro\meter}$, the heat flux is smaller than that of the black-body radiation. This agrees with the fact that heat transfer is mainly achieved by propagating waves in the far field \cite{song}.

\begin{figure}
	\includegraphics[width=8.6 cm]{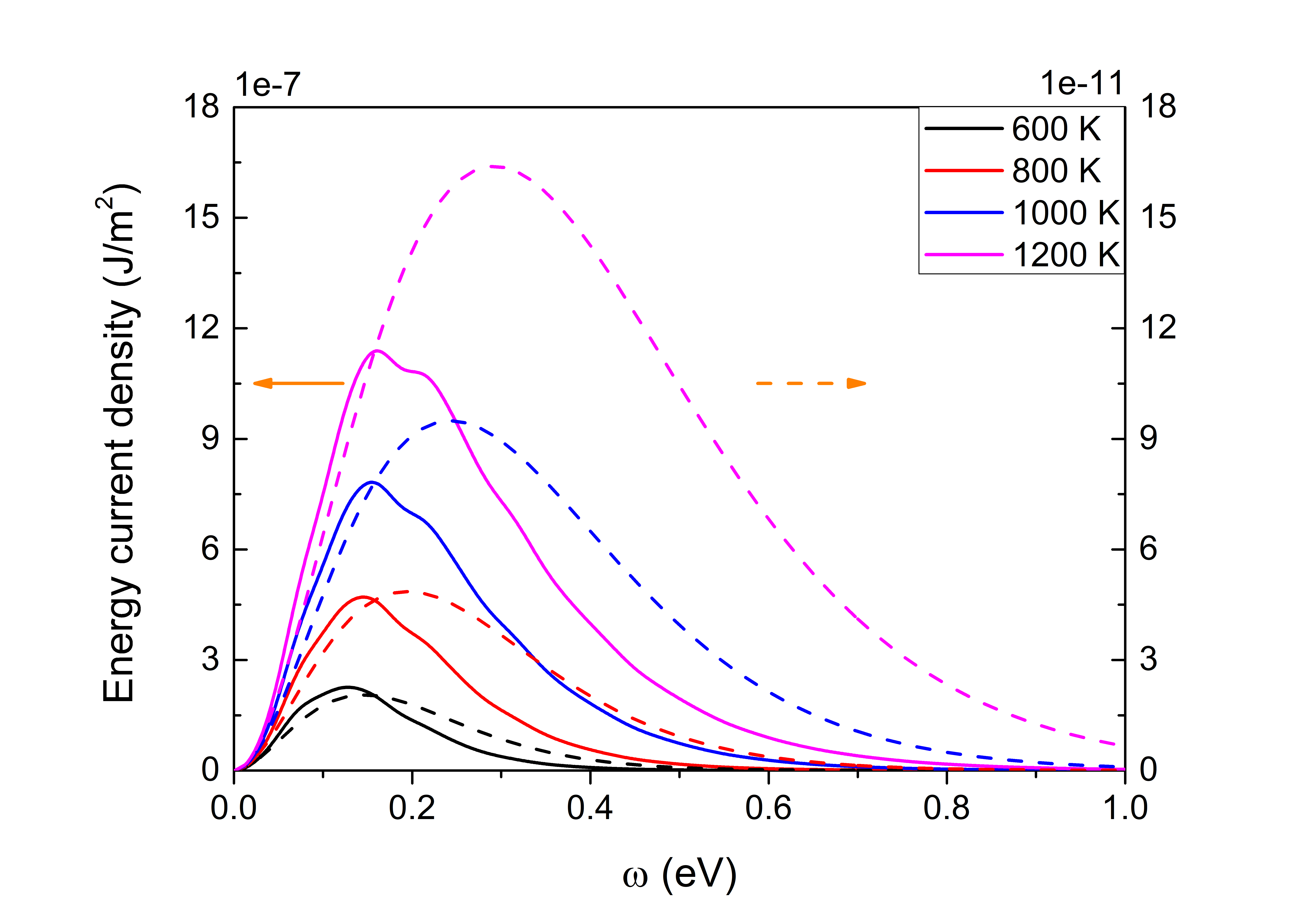}
	\caption{The Coulomb (solid curve, left axis, $d=100$ nm) and black-body radiative (dashed curve, right axis) energy current density of graphene with different temperatures. The left and right axes have the same scale of ticks but differ by 4 orders of magnitude.}
	\label{fig3}
\end{figure}

Figure \ref{fig3} compares energy current densities obtained from the \textit{ab initio} calculation of the Coulomb-type heat transfer with Plank's law of black-body radiation. The black-body energy current density depends on only frequency and temperature, while the Coulomb interaction decays exponentially with the gap sizes, as shown in Eq.~(\ref{coulomb2d}). To eliminate the non-local effects, we calculate the Coulomb energy current density between two graphene sheets at the separation $d=$ 100 nm. As shown in Fig.~\ref{fig3}, the Coulomb energy current density is approximately 4 orders of magnitude larger than that of black-body radiation. Moreover, most spectrum weights of Coulomb energy current density lie in the low frequencies, while the black-body radiation has a broader spectrum. When $\omega<\SI{0.16}{\electronvolt}$, the Coulomb energy current density has an increasing trend similar to that of the black-body spectrum. However, with a further increase in frequency, the Coulomb energy current density decays rapidly and vanishes at high frequencies. This implies that the near-field enhancement of the Coulomb energy current between two graphene sheets is mainly attributed to electron transitions around the Fermi level.

\begin{figure}
	\includegraphics[width=8.6 cm]{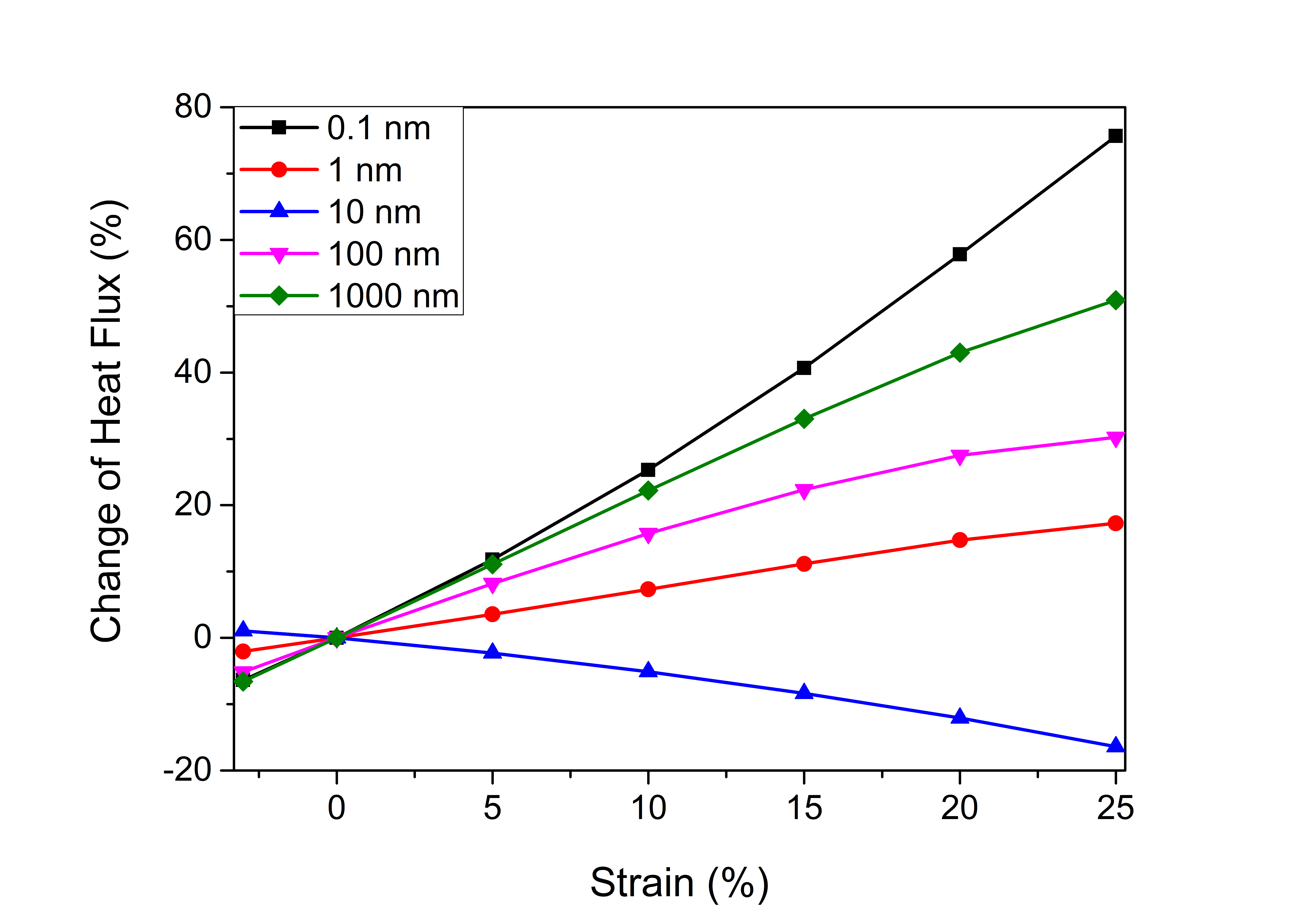}
	\caption{Strain effect on the Coulomb heat flux between two parallel single-layer graphene sheets with different gap sizes. The horizontal axis is the magnitude of strain in percent, and the vertical axis is the percentage change of heat flux after strain is applied.}
	\label{fig4}
\end{figure}

Another advantage of our method is that one can easily change the parameters of materials without further theoretical treatment of models. Experimentally, the strain arises naturally due to the surface corrugations \cite{strain1} or lattice mismatch between graphene and the substrate \cite{strain2}. In Fig.~\ref{fig4}, we show the strain effect on the heat flux between two graphene sheets with different gap sizes. We introduce the uniform biaxial strain by varying the lattice constant with a percentage change from $-3\%$ to $25\%$ and calculate the heat flux after strain is applied. As shown in Fig.~\ref{fig4}, the heat flux is positively correlated to the strain for most of the gap size. This enhancement may originate from effective electronic scalar potential induced by uniform strain in graphene \cite{strain3,strain4}. Interestingly, with an increase in gap size, the enhancement becomes less significant at $d=1$ nm, and even a negative dependence is shown at $d=10$ nm. This may be due to the interplay between local and nonlocal effects in calculating the transmission coefficient. At short distances, the heat flux is calculated with full nonlocal effects. However, with the increase in the distances, finite wave vectors are gradually killed by the exponential factor $e^{-qd}$ in the Coulomb potential, so that only partial finite wave vectors contribute at moderate distances. With a further increase of the gap size, the long-wave component dominates, and the heat flux shows a positive dependence again. Similar behavior was reported in Ref.~\cite{strain5}, where such unusual strain dependence may be related to the bandwidth of the spectral transmission coefficient. However, a decisive quantitative analysis of this phenomenon is still lacking, and we leave it for future works.

{\emph{Summary.}}
In summary, we have presented a general first-principles method to study the near-field heat transfer mediated by Coulomb interaction. Using the NEGF formalism, we derived Caroli formulas for both 2D and 3D materials with translational invariance. Within the density functional theory, we calculated the polarizability from electronic band structures and obtained the screened Coulomb interaction with the retarded Dyson equation. With the macroscopic approximation, we showed that the Caroli formula for 2D materials reduces to the $p$-polarized evanescent modes in the FE theory. We adopted single-layer graphene as an example to show the procedures for calculating the heat flux between two parallel sheets. Our results show that the near-field Coulomb heat flux is significantly larger than that of the black-body radiation, and an asymptotic dependence of 1/$d^2$ was shown at large separations. With spatial dispersion, the heat flux saturates at small distances, while $1/d$ dependence was shown for results obtained from the local response function. The spectrum weight of the Coulomb energy current concentrates at low frequencies, which implies that the near-field enhancement originates from electron transitions around the Fermi level. For most of the gap sizes, the calculated heat flux has a positive correlation with the strain applied. A negative correlation is shown at moderate distances which may relate to the interplay between local and nonlocal effects or to the narrow bandwidth of the spectral transmission coefficient. Our method can be used to study the Coulomb heat transfer of a wide range of materials with finite thickness and possible inhomogeneities, which provides a benchmarking reference for both theory and experiment.         

{\emph{Acknowledgments.}}
%\section*{Acknowledgments}
We would like to thank J. Peng for discussions on fluctuational electrodynamics. This work is supported by FRC Grant No. R-144-000-427-114 and MOE tier 2 Grant No. R-144-000-411-112.
%\vfill


\begin{thebibliography}{[Vo]}
	\bibitem{volokitin}
	A. I. Volokitin and B. N. J. Persson, Rev. Mod. Phys. \textbf{79}, 1291 (2007).
	\bibitem{zzm}
	Z. M. Zhang, \textit{Nano/Microscale Heat Transfer} (Springer, Cham, 2020).
	\bibitem{biehs}
	S.-A. Biehs, R. Messina, P. S. Venkataram, A. W. Rodriguez, J. C. Cuevas, P. Ben-Abdallah, Rev. Mod. Phys. \textbf{93}, 025009 (2021).
	\bibitem{experiment1}
	C. M. Hargreaves, Phys. Lett. A \textbf{30}, 491 (1969).
	\bibitem{experiment2}
	G. A. Domoto, R. F. Boehm, and C. L. Tien, J. Heat Transfer. \textbf{92}(3), 412 (1970).
	\bibitem{experiment3}
	R. S. Ottens, V. Quetschke, S. Wise, A. A. Alemi, R. Lundock, G. Mueller, D. H. Reitze, D. B. Tanner, and B. F. Whiting, Phys. Rev. Lett. \textbf{107} 014301 (2011).
	\bibitem{rytov1}
	S. M. Rytov, \textit{Theory of electric fluctuations and thermal radiation} (Air Force Cambridge Research Center, Bedford, MA, 1953).
	\bibitem{rytov2}
	S. M. Rytov, Y. A. Kravtsov, and V. I. Tatarskii, \textit{Principles of statistical radiophysics} (Springer, Berlin, 1989).
	\bibitem{polder}
	D. Polder and M. A. Van Hove, Phys. Rev. B \textbf{4}, 3303 (1971).
	\bibitem{fdt1}
	H. Nyquist, Phys. Rev. \textbf{32} 110 (1928). 
	\bibitem{fdt2} 
	H. B. Callen and T. A. Welton, Phys. Rev. \textbf{83} 34 (1951).
	\bibitem{song}
	B. Song, A. Fiorino, E. Meyhofer, and P. Reddy, AIP Adv. \textbf{5}, 053503 (2015).
	\bibitem{cuevas}
	J. C. Cuevas, and F. J. Garc\'ia-Vidal, ACS Photonics \textbf{5}, 3896 (2018).
	\bibitem{experiment4}
	A. Kittel, W. M\:uller-Hirsch, J. Parisi, S. -A. Biehs, D. Reddig, and M. Holthaus, Phys. Rev. Lett. \textbf{95}, 224301 (2005).
	\bibitem{experiment5}
	I. Altfeder, A. A. Voevodin, and A. K. Roy, Phys. Rev. Lett. \textbf{105}, 166101 (2010).
	\bibitem{experiment6}
	K. Kloppstech, N. Könne, S.-A. Biehs, A. W. Rodriguez, L. Worbes, D. Hellmann, and A. Kittel, Nat. Commun. \textbf{8}, 14475 (2017).
	\bibitem{experiment7}
	L. Worbes, D. Hellmann, and A. Kittel, Phys. Rev. Lett. \textbf{110}, 134302 (2013).
	\bibitem{vazrik}
	V. Chiloyan, J. Garg, K. Esfarjani and G. Chen, Nat. Commun. \textbf{6}, 6755 (2015).
	\bibitem{xiong}
	S. Xiong, K. Yang, Y. A. Kosevich, Y. Chalopin, R. D'Agosta, P. Cortona, and S. Volz, Phys. Rev. Lett. \textbf{112}, 114301 (2014). 
	\bibitem{cui}
	L. Cui, W. Jeong, V. Fern\'andez-Hurtado, J. Feist, F. J. Garc\'ia-Vidal, J. C. Cuevas, E. Meyhofer, and P. Reddy, Nat. Commun. \textbf{8}, 14479 (2017).
    \bibitem{lfe1}
    S. G. Louie, J. R. Chelikowsky, and M. L. Cohen, Phys. Rev. Lett. \textbf{34}, 155 (1975).
	\bibitem{lfe2}
	T. Zhu, P. E. Trevisanutto, T. C. Asmara, L. Xu, Y. P. Feng, and A. Rusydi, Phys. Rev. B \textbf{98}, 235115 (2018). 
    \bibitem{negf1}
	J.-S. Wang and J. Peng, arXiv:1607.02840.
	\bibitem{negf2}
	J.-S. Wang and J. Peng, EPL, \textbf{118} 24001 (2017).
	\bibitem{negf3}
	J. Peng, H. H. Yap, G. Zhang, and J.-S. Wang, arXiv:1703.07113.
	\bibitem{negf4}
	G. Tang and J.-S. Wang, Phys. Rev. B \textbf{98}, 125401 (2018).
	\bibitem{mahan} 
	G. D. Mahan, Phys. Rev. B \textbf{95}, 115427 (2017). 
	\bibitem{zhang}
	Z.-Q. Zhang, J. T. L\"u, and J. -S. Wang, Phys. Rev. B \textbf{97}, 195450 (2018). 
	\bibitem{caroli}
	C. Caroli, R. Combescot, P. Nozieres, and D. Saint-James, J. Phys. C \textbf{4}, 916 (1971).
	\bibitem{Yu} 
	R. Yu, A. Manjavacas, and F. J. Garc\'ia de Abajo, Nat. Commun. \textbf{8}, 2 (2017).
	\bibitem{equivalent} 
	J.-S. Wang, Z.-Q. Zhang, and J. T. L\"u, Phys. Rev. E \textbf{98}, 012118 (2018).
	\bibitem{zhu}
	T. Zhu, Z.-Q. Zhang, Z. Gao, and J.-S. Wang, Phys. Rev. Appl. \textbf{14} 024080 (2020).
	\bibitem{jiang}
	J.-H. Jiang, and J.-S.Wang, Phys. Rev. B \textbf{96}, 155437 (2017).
	\bibitem{tang}
	G. Tang, H. H. Yap, J. Ren, and J.-S. Wang, Phys. Rev. Applied \textbf{11}, 031004(R) (2019).
	\bibitem{lian}
	K. N. Lian and J.-S. Wang, Eur. Phys. J. B \textbf{93}, 138 (2020). 
	\bibitem{Onida}
	G. Onida, L. Reining, and A. Rubio, Rev. Mod. Phys. \textbf{74}, 601 (2002).
	\bibitem{meir1}
	Y. Meir and N. S. Wingreen, Phys. Rev. Lett. \textbf{68}, 2512 (1992).
	\bibitem{meir2}
	A.-P. Jauho, N. S.Wingreen, and Y. Meir, Phys. Rev. B \textbf{50}, 5528 (1994).
	\bibitem{dyson}
	F. J. Dyson, Phys. Rev. \textbf{75}, 1736 (1949).
	\bibitem{rpa}
	H. Ehrenreich, and M. H. Cohen, Phys. Rev. \textbf{115}, 786 (1959).
	\bibitem{kubo}
	R. Kubo, J. Phys. Soc. Jpn. \textbf{12}, 570 (1957); R. Kubo, M. Yokota, and S. Nakajima, \textit{ibid.} \textbf{12}, 1203 (1957).
	\bibitem{polarizability}
	T. Zhu, M. Antezza and J.-S. Wang, Phys. Rev. B \textbf{103}, 125421 (2021).
	\bibitem{eta}
	N. D. Mermin, Phys. Rev. B \textbf{1}, 2362 (1970).
	\bibitem{adler}
	S. L. Adler, Phys. Rev. \textbf{126}, 413 (1962).
	\bibitem{wiser}
	N. Wiser, Phys. Rev. \textbf{129}, 62 (1963).
	\bibitem{ks}
	W. Kohn, and L. Sham, Phys. Rev. \textbf{140}, A1133 (1965).
	\bibitem{quek}
	F. Xuan, Y. Chen, and S. Y. Quek, J. Chem. Theory Comput. \textbf{15}, 3824 (2019).
	\bibitem{2dpi}
	F. A. Rasmussen, Ph.D. dissertation, Technical University of Denmark, 2016.
	\bibitem{ilic}
	O. Ilic, M. Jablan, J. D. Joannopoulos, I. Celanovic, H. Buljan, and M. Solja\v{c}i\'{c}, Phys. Rev. B \textbf{85}, 155422 (2012).
	\bibitem{ilic2}
	O. Ilic, M. Jablan, J. D. Joannopoulos, I. Celanovic, and M. Solja\v{c}i\'{c}, Opt. Express \textbf{20}, A366 (2012).
	\bibitem{pablo}
	P. R. -L\'opez, W. -K. Tse, and D. A. R. Dalvit, J. Phys.: Condens. Matter \textbf{27}, 214019 (2015).
	\bibitem{qe1}
	P. Giannozzi, S. Baroni, N. Bonini, M. Calandra, R. Car, C. Cavazzoni, D. Ceresoli, G. L. Chiarotti, M. Cococcioni, I. Dabo \textit{et al.}, J.Phys.:Condens.Matter \textbf{21}, 395502 (2009).
	\bibitem{qe2}
	P. Giannozzi, O. Andreussi, T. Brumme, O. Bunau, M. B. Nardelli, M. Calandra, R. Car, C. Cavazzoni, D. Ceresoli, M. Cococcioni \textit{et al.}, J.Phys.:Condens.Matter \textbf{29}, 465901 (2017).
	\bibitem{martin}
	N. Troullier and J. L. Martins, Phys. Rev. B \textbf{43}, 1993 (1991).
	\bibitem{pbe}
	J. P. Perdew, K. Burke, and M. Ernzerhof, Phys. Rev. Lett. \textbf{77}, 3865 (1996).
	\bibitem{bgw1}
	M. S. Hybertsen and S. G. Louie, Phys. Rev. B \textbf{34}, 5390 (1986).
	\bibitem{bgw2}
	J. Deslippe, G. Samsonidze, D. A. Strubbe, M. Jain, M. L. Cohen, and S. G. Louie, Comput. Phys. Commun. \textbf{183}, 1269 (2012).
	\bibitem{mp}
	H. J. Monkhorst, and J. D. Pack, Phys. Rev. B \textbf{13}, 5188 (1976).
	\bibitem{nonlocal}
	P. -O. Chapuis, S. Volz, C. Henkel, K. Joulain, and J. -J. Greffet, Phys. Rev. B \textbf{77}, 035431 (2008).
	\bibitem{grapheneexp}
	J. Yang, W. Du, Y. Su, Y. Fu, S. Gong, S. He, and Y. Ma, Nat Commun \textbf{9}, 4033 (2018).
	\bibitem{strain1}
	M. Teague, A. Lai, J. Velasco, C. Hughes, A. Beyer, M. Bockrath, C. Lau and N.-C. Yeh, Nano Lett. \textbf{9}, 2542 (2009).
	\bibitem{strain2}
	Z. H. Ni, W. Chen, X. F. Fan, J. L. Kuo, T. Yu, A. T. S. Wee, and Z. X. Shen, Phys. Rev. B \textbf{77}, 115416 (2008).
	\bibitem{strain3}
	S.-M. Choi, S.-H. Jhi, and Y.-W. Son, Phys. Rev. B \textbf{81}, 081407(R) (2010).
	\bibitem{strain4}
	C. Si, Z. Sun, and F. Liu, Nanoscale \textbf{8}, 3207 (2016).
	\bibitem{strain5}
	L. Ge, Z. Xu, Y. Cang, and K. Gong, Opt. Express \textbf{27}, A1109 (2019). 
\end{thebibliography}
\end{document}